\documentclass[aps,prl,preprint,]{revtex4}

\newcommand{\beq}{\begin{equation}}
\newcommand{\eeq}{\end{equation}}
\newcommand{\bea}{\begin{eqnarray}}
\newcommand{\eea}{\end{eqnarray}}

\begin{document}

\title{ROSENFELD, BERGMANN, DIRAC AND THE INVENTION OF CONSTRAINED HAMILTONIAN DYNAMICS}

\author{D. C. SALISBURY}

\affiliation{DEPARTMENT OF PHYSICS, AUSTIN COLLEGE,
Sherman, TX  75090, USA\\E-mail:
dsalisbury@austincollege.edu}

\begin{abstract}
In a paper appearing in Annalen der Physik in 1930 Leon Rosenfeld invented the first  procedure for producing Hamiltonian constraints. He displayed and correctly distinguished the vanishing Hamiltonian generator of time evolution, and the vanishing generator of gauge transformations for general relativity with Dirac electron and electrodynamic field sources. Though he did not do so, had he chosen one of his tetrad fields to be normal to his spacetime foliation, he would have anticipated by almost thirty years the general relativisitic Hamiltonian first published by Paul Dirac. 
\end{abstract}

\maketitle

\section{Introduction and obstacles to quantizing electrodynamics}

Leon Rosenfeld produced his groundbreaking constrained Hamiltonian dynamics formalism, published in Annalen der Physik in 1930 under the title {\it On the Quantization of Wave Fields}\cite{rosenfeld30}, in those heady times shortly after Dirac had achieved his relativistic quantum theory of the electron. Heisenberg and Pauli were quantizing the electromagnetic field while Weyl and Fock had shown how to couple Dirac's electron field to gravity. A fundamental unification seemed imminent. The confident young Rosenfeld, inspired by his mentor Wolfgang Pauli, proposed precisely a quantum field theoretic unification of  gravity and electromagnetism. And he came surprisingly close! Sadly it appears that neither he nor Peter Bergmann or Paul Dirac, both of whom  began nearly twenty years later to address the problem of converting singular Lagrangian systems into Hamiltonian models, fully appreciated the enormous progress he made in his 1930 paper.  I will sketch in this short article only some aspects of Rosenfeld's analysis, with an effort to highlight contributions that were independently  reinvented decades later. A full translation of Rosenfeld's work with commentary will appear elsewhere.

Emmy Noether showed in 1918 that if a dynamical model possesses a symmetry under a  transformations involving arbitrary functions then a specific linear combination of equations of motions must vanish identically \cite{noether18}. Thus, for example,  the Bianchi identity in general relativity is a reflection and consequence of the general covariance of Einstein's equations. Similarly, since classical electrodynamics is covariant under the gauge transformation of the electromagnetic four-potential $A_\mu$, where the transformed potential is $A'_\mu = A_\mu + \partial_\mu \Lambda$ and $\Lambda$ is an arbitrary spacetime function, then Noether's theorem shows that $F^{\mu \nu}_{,\mu \nu}$ must vanish identically, where $F^{\mu \nu}$ is the electromagnetic field tensor. Related to this symmetry is the vanishing of the momentum associated with the naught component of the potential. This posed a problem for researchers attempting to quantize the electromagnetic field in the late 1920's. Heisenberg and Pauli had proposed two not entirely satisfactory methods for dealing with this embarrassment. These procedures destroyed either manifest gauge or manifest Lorentz symmetry. Pauli is quoted having said ``Ich warne Neugierige",``I forewarn the curious". Rosenfeld was in 1929 collaborating with Pauli in Zurich, and it was Pauli who encouraged him to construct a general manifestly symmetric formalism. Rosenfeld writes in the 1930 article (my translation) ``As I was investigating these relations in the especially instructive example of gravitation theory, Professor Pauli helpfully indicated to me the principles of a simpler and more natural manner of applying the Hamiltionian procedure in the presence of identities".  

Setting equal to zero coefficients of the highest time derivatives of the arbitrary gauge functions in Noether's identities, Rosenfeld discovered three interrelated consequences:
\begin{itemize}
\item There are as many primary constraints, i.e., identically vanishing functions of configuration variables and momenta (conceived as functions of configuration and velocity), as there are arbitrary gauge functions.
 
 \item The Legendre matrix, consisting of second partial derivatives of the Lagrangian with respect to velocities, is singular.
 
 \item Rosenfeld considered only Lagrangians quadratic in velocities. Consequently all momenta involved contractions of the singular Legendre matrix with velocities. Therefore it was possible to add arbitrary linear combinations of null vectors to velocities without altering the momenta. These linear combinations reflect the arbitrariness in evolution in time of initial data.
\end{itemize}
All of these results were obtained independently by Bergmann in 1949.\cite{bergmann49a} 

Rosenfeld then supposed that solutions had been found for all velocities in terms of momenta, including admissible arbitrary functions, and he constructed a Hamiltonian with the canonical expression augmented by additional linear combinations of primary constraints. Bergmann and Brunings obtained a similar formal result in 1949.\cite{bergbrun49} Bergmann, Schiller, and Zatkis in 1950 invented an algorithm for solving for the velocities in terms of the momenta.\cite{bergPSZ50} Rosenfeld never addressed this general question. In 1949 Dirac approached the construction of the Hamiltonian for singular systems from an entirely different perspective.\cite{dirac50} His work was first published in 1950. He was motivated by a desire to choose arbitrary time foliations in flat spacetime. Dirac never concerned himself with the faithful  reproduction in the canonical Hamiltonian framework of Lagrangian symmetries. This was a principle focus of both Rosenfeld and Bergmann.

Indeed, Rosenfeld found the correct form for canonical generators of gauge symmetries, expressed as a sum of geometric part (determined by the tensorial nature of the variables undergoing variations, and a transport term (reflecting the fact that active variations were contemplated at a fixed coordinate location). He proved that his generator produced the correct variation not only of configuration but also of momentum variables. And in a culminating {\it tour de force} he proved that while the symmetry generator contained the primary constraints multiplying the highest time derivatives of the gauge functions, the preservation in time of the entire generator implied that the coefficients of all lower time derivatives of the gauge functions must themselves be constraints. In other words, Rosenfeld was the original inventor of the what is now referred to as the "Dirac-Bergmann" algorithmn!  Indeed, the Rosenfeld analysis yielded all constraints in a single step, a perspective that conflicts with the terms ``primary", ``secondary", etc. first introduced in 1951 by Anderson and Bergmann  to characterize constraints.\cite{bergA51}

\section{The Hamiltonian formulation of general relativity}

Rosenfeld came surprisingly close to the breakthrough published by Dirac in 1958\cite{dirac58}, and discovered independently at about the same time by B.  DeWitt (unpublished) and Anderson\cite{anderson58}. Dirac showed that through subtraction of an appropriate total derivative from the Weyl gravitational Lagrangian that time derivatives of the naught components of the metric could be eliminated, resulting in trivially vanishing conjugate momenta. Weyl removed second derivatives of the metric by eliminating derivatives of the Christofel tensor through the subtraction of an appropriate total divergence.\cite{weyl18}  Rosenfeld considered a tetrad form of gravity. Similarly to Weyl, he eliminated second derivatives of the tetrad fields by removing derivatives of the Ricci rotation coefficients through the subtraction from the Hilbert action of an appropriate total divergence. It turns out that  if he had simply adapted his tetrad to his spacetime foliation by taking one of the orthonormal tetrad vectors to point perpendicular to the fixed time hypersurfaces while the remaining triads were tangent to the foliation, he would have obtained a Lagrangian in which no time derivatives of the orthonormal tetrad components appear. Thus he would have anticipated Dirac, Anderson, and DeWitt by almost three decades. Had he expressed this orthonormal tetrad in terms of the lapse and shift functions introduced by Arnowitt, Deser and Misner he would have obtained the triad form of their ADM Hamiltonian.\cite{adm62}

\end{document}